# A new approach for an unitary risk theory

NICOLAE POPOVICIU
Hyperion University of Bucharest, Faculty of Mathematics-Informatics
Street Călăraşilor 169, Bucharest, ROMANIA
FLOAREA BAICU
Hyperion University of Bucharest, Faculty of Mathematics-Informatics
Street Călăraşilor 169, Bucharest, ROMANIA
nic.popoviciu@yahoo.com ; floareabaicu@zappmobile.ro

*Abstract*. The work deals with the risk assessment theory. An unitary risk algorithm is elaborated. The algorithm is based on parallel curves. The basic curve of risk is a hyperbolic curve, obtained as a multiplication between the probability of occurrence of certain event and its impact. Section 1 contains the problem formulation. Section 2 contains some specific notations and the mathematical background of risk algorithm. A numerical application based on risk algorithm is the content of section 3. Section 4 contains several conclusions.

## 1. Introduction. Problem statement

The topic of the work is the following: notations and specific definitions, the mathematical background of risk theory and the risk computation algorithm, followed by a numerical example.

### 1.1. Specific definitions

**The risk** is defined by the international standard ISO/CEI 17799 [1] as being the combination between the probability of occurrence of certain event and its consequences.

**The risk level** is an arbitrary indicator, denoted L, which allows grouping certain risks into equivalency classes. These classes include risks which are placed between two limit levels – acceptable and unacceptable – conventionally established.

**The acceptable risk level** is the risk level conventionally admitted by the organization management, regarded as not causing undesirable impacts on its activity. This is determined by methodic and specific evaluations.

**The residual risk** is considered to be the reminder after the risk treatment. As a general rule, the residual risk may be regarded as risk on acceptable level.

Assuming the risk definition set forth upwards, this is a positive real number $R$ which may be represented by the area of a quadrangle surface, having as one side the **P**robability of occurrence of certain event, noted with $P$, and as the other side the consequences of that event occurrence, respectively the **I**mpact of the undesirable event, noted with $I$, upon the security of the studied organization.

Mathematically speaking, the same area may be obtained through various combinations between $P$ and $I$, of which the preferred one is quite the product between probability and impact.

There are a lot of **P**robability – **I**mpact couples generating the same **R**isk $R$, defining quadrangles of same area as illustrated in figure 1.

If the vertexes of such quadrangles, which are not on axes, are linked through a continuous line it results a **hyperbolic curve** $C$, named the **curve of risk** [2]. This curve allows the differentiation between the acceptable risk (Tolerable – T) and the unacceptable one (Non-Tolerable – NT).

Thus, the risk of occurrence of a certain event A, with high impact, with serious consequences but low probability of occurrence, defined by coordinates placed below the represented acceptability curve is considered **acceptable**, while the risk of event B, with less serious consequences but high probability of occurrence, of which coordinates are placed upwards the curve, is considered **unacceptable**.

Hyperbolic curve of risk based on couples (**P**robability, **I**mpact) is illustrated in figure 1.

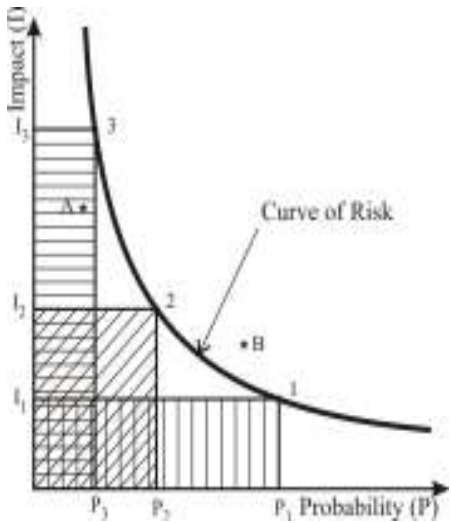

**Figure 1.** Graph representation for equivalency of risks defined by different probability – impact couples

Any placement of an assessed risk on a particulate risk level L represents a major and critical decision for any manager of security system and raises minimum three heavy and sensitive problems, respectively [2,3,4]:

1. How to define the risk coordinates, respectively the Probability – Impact couple?

2. How to set out the Curve of Risk for assessing which risk is tolerable (acceptable) and which is not?

3. How many curves are to be set out, respectively how many risk levels are to be defined and which is the **distance** h between different curves?.

**1.2. Classes of impact and probability**

A class of 7 levels $I_1, I_2, \cdots, I_7$ is proposed in relation to impact, respectively: insignificant, very low, low, medium, high, very high and critical. These impact classes are established considering the values of losses that might occur as a consequence of a risk [2].

As to the probability of event occurrence the proposal is for 9 (nine) classes, by assigning a *class of probability* to each probability of occurrence, starting with the most rare event, class 1, and up to the most frequent event, the class counted to be the highest one, class 9 like is presented in the table 1, column 3.

For the lowest probability of occurrence selected for this work, class of probability 1, the assumption for the occurrence of a certain event is once at every 20 years, respectively an event at 175200 hours, gives a probability of $1/175200 \approx 5 \cdot 10^{-6}$, mentioned in column 2, line 1 of table 1.

For the second probability of occurrence mentioned in column 3 of table 1, the assumption for the occurrence of a certain event is once at every 2 years, respectively a probability of $1/17520 \approx 5 \cdot 10^{-5}$.

For all the other probabilities mentioned in column 2 of table 1, the assumption is of more frequent occurrences, namely those mentioned in column 1 of this table. The calculation is made similarly, the probabilities being based on the occurrence frequency measured in hours [2], [3].

It is assumed that the most frequent event occurs during each functioning hour, its probability being 1, line 9 of table 1.

Table 1 with the classes of probabilities.

| No. | Frequency of occurrence | Probability | Class of probability |
|---|---|---|---|
| 0 | 1 | 2 | 3 |
| 1 | One event at 20 years | $5 \cdot 10^{-6}$ | 1 |
| 2 | One event at 2 years | $5 \cdot 10^{-5}$ | 2 |
| 3 | One event per year | $10^{-4}$ | 3 |
| 4 | One event at 6 months | $2 \cdot 10^{-4}$ | 4 |
| 5 | One event per month | $1.4 \cdot 10^{-3}$ | 5 |
| 6 | Two events per month | $3 \cdot 10^{-3}$ | 6 |
| 7 | One event per week | $6 \cdot 10^{-3}$ h$^{-}$ | 7 |
| 8 | One event per day | $4 \cdot 10^{-2}$ h$^{-1}$ | 8 |
| 9 | One event at each hour | 1 h$^{-1}$ | 9 |

**1.3 Risk levels**

The sections delimited between two consecutive curves $C_j$ represent the levels of risk. We consider the case $(C_j, C_{j+1}]$.

**2. Mathematical background**

In an orthogonal system $xOy$ we use a set of **known points:**

$A_{i,j} = A(x_i, y_j), i = 1, m; j = 1, n$

$(x_i, y_j) \in N \times N$ or $(x_i, y_j) \in R^+ \times R^+$ where

$m$ = number of the probability classes;
$n$ = number of the impact classes.

Explicitly, the points $A_{i,j}$ are

$(x_1, y_1)(x_2, y_1) \cdots (x_m, y_1)$ impact of level 1;
$(x_1, y_2)(x_2, y_2) \cdots (x_m, y_2)$ impact of level 2;
…………………………………..
$(x_1, y_n)(x_2, y_n) \cdots (x_m, y_n)$ impact of level n.

We propose that the point $(x_m, y_1)$ belongs to the hyperbolic curve

$C_1: xy = R_1$, where $R_1 = x_m y_1$ is the **risk value** (or risk measure, measure of risk) and so $C_1$ is the **curve of risk** for the risk value $R_1$.

Let $r$ be a natural number which represents the total number of curves of risk. The value of $r$ is at user's disposal. Hence there exists the risk curves $C_1, \cdots, C_r$.

Now the problem is: how to construct them? There exists several possibilities. In this work we chose the case of **parallel curves**.

## 2.1. The parallel curves

A local problem formulation is the following: given a known curve $\gamma$, let us construct a new curve $\Gamma$ so that $\gamma // \Gamma$ ($\Gamma$ parallel with $\gamma$). In our case $\gamma$ is hyperbolic curve. Let $(x, y)$ be a current point on $\gamma$ and $(X, Y)$ a current point on $\Gamma$.

*Proposition 1*. If $\gamma$ has the explicit equation $y = \dfrac{c}{x}$, $x \neq 0$, $c > 0$ then the parallel curve $\Gamma$ has the **parametric equations:**

$$X = x + \frac{ch}{\sqrt{c^2 + x^4}}, \quad Y = \frac{c}{x} + \frac{hx^2}{\sqrt{c^2 + x^4}} \quad (1)$$

where $h = d(M, N)$, $M \in \gamma$, $N \in \Gamma$ is the constant distance between any two points on the common normal.

*Proof.* For the moment we denote $M(x_0, y_0)$, $x_0 y_0 = c$. The normal in the point $M \in \gamma$ has the equation $y - y_0 = \dfrac{x_0^2}{c}(x - x_0)$. (2)

We denote $X = x_0 + p, Y = y_0 + q$, $p^2 + q^2 = h^2$ and put the condition that $N(X, Y)$ verifies the equation (2). So we obtain $q = \dfrac{x_0^2}{c} p$ and then $p = \pm \dfrac{ch}{\sqrt{c^2 + x_0^4}}$.

We took $h > 0$, $p > 0$, $q > 0$ and finally we obtain

$$p = \frac{ch}{\sqrt{c^2 + x_0^4}}, \quad q = \frac{hx_0^2}{\sqrt{c^2 + x_0^4}}. \quad (3)$$

Because $x_0$ and $y_0$ have been arbitrary values, one obtains the relations (1). (End).

*Remark 1*. It is very difficult to eliminate the variable $x$ and to obtain the explicit equation $Y = F(X)$ for the curve $\Gamma$. In numerical applications we have $c = R_1$.

In order to draw the parallel curves $\gamma$ and $\Gamma$, let us say by a Mathcad Professional subroutine, we use the functions [5]:

$$y = \frac{c}{x}; \quad X(x, c, h) = x + \frac{ch}{\sqrt{c^2 + x^4}}; \quad (4)$$

$$Y(x, c, h) = \frac{c}{x} + \frac{hx^2}{\sqrt{c^2 + x^4}}.$$

and apply them by Insert, Graph, X-Y Plot.

*Remark 2*. Also it is possible to formulate the inverse problem: having the equation of $\gamma$ and the known point $N(a, b)$, we look for the value of $h$ and the equation of $\Gamma$ so that $N \in \Gamma$ and $\Gamma // \gamma$.

*Proposition 2*. In the above conditions we obtain the equation:

$$x^4 - ax^3 + bcx - c^2 = 0, \quad a, b, c \text{ known} \quad (5)$$

*Proof.* One eliminates $h$ from the parametric equations

$$x + \frac{hc}{\sqrt{c^2 + x^4}} = a, \quad \frac{c}{x} + \frac{hx^2}{\sqrt{c^2 + x^4}} = b$$

(End).

By a numerical method we solve the equation (5) and let $\bar{x}$ be the chosen solution. Then

$$h = \frac{a - \bar{x}}{c}\sqrt{c^2 + \bar{x}^4}.$$

Having the value $h$ we can construct $\Gamma$ by (4).

## 2.2. The construction of r parallel curves of risk. The risk algorithm.

We recall that the parameter $r$ defines the total number of parallel curves of risk and $r$ is known i.e. it is given by the user. So we have to construct $r$ parallel curves of risk $C_j, j = 1, r$ where $C_1$ is known. The others curves $C_2, \cdots, C_r$ are constructed in the manner shown below and called the **algorithm p. c.** ( algorithm of parallel curves ).

1. We denote $V$ the point $A(x_m, y_n)$ and the equation of line $(OV)$ is $y = (y_n x)/x_m$.

2. Let $B_1$ be the point defined as $B_1 = (OV) \cap C_1$ and denote $B_1(a_1, b_1)$.

**3**. Write the normal in $B_1$ for the hyperbolic curve $f(x) = c/x$, $x \neq 0$. We obtain successively the slopes and the equation:

$$m_{\tan} = -\frac{c}{a_1^2}, \quad m_{norm} = \frac{a_1^2}{c}$$

$y - b_1 = m_{norm}(x - a_1)$ (the normal in $B_1$).

**4**. Let $B_{r+1}$ be the point defined as the intersection
$B_{r+1} = (normal\ in\ B_1) \cap (y = y_n)$
$B_{r+1}(a_{r+1}, b_{r+1})$.

**5**. Compute the distance $\|B_1 B_{r+1}\|$ and denote the step between the parallel curves by $h = \|B_1 B_{r+1}\|/r$, where $h$ appears in parametric equations (1).

**6**. We use the following correspondence for curves
$C_2$ with $h_2 = h$, $C_3$ with $h_3 = 2h$ and so on
$C_r$ with $h_r = (r-1)h$       (6)

The values $h_j$, $j = 2, r$ are used in equations (4).

**7**. Compute the risk values $R_j$ corresponding to each curve $C_j$, $j = 2, r$. Up to now we know, by computation, the coordinates of the points $B_1$ and $B_{r+1}$.

7.1. We need the coordinates of the points $B_2(a_2, b_2), \cdots, B_r(a_r, b_r)$.
For that one uses the formulas

$$\frac{\|B_1 B_j\|}{\|B_j B_{r+1}\|} = k_j, \quad j = 2, r, \quad k_j \in Q$$

$$a_j = \frac{a_1 + k_j a_{r+1}}{1 + k_j}, \quad b_j = \frac{b_1 + k_j b_{r+1}}{1 + k_j}, \quad j = 2, r$$

7.2. The risk values $R_j$ are $R_j = a_j b_j$ and the parallel curves equations are $(C_j): xy = R_j$, $j = 2, r$ or are given by equivalent parametric equations (4).

**8**. We draw on the same orthogonal system $xOy$ all the parallel curves $C_j$, $j = 1, r$ and all the points $A_{ij}$ i.e. all the lines
$x = x_i$, $i = 1, m$; $y = y_j$, $j = 1, n$

We remark that the space $R^2$ could be considered a affined space in this drawing.

**9**. The **level of risk** is defined as the set of points $A_{ij}$ settled between two successive curves of risk. We consider the cases $(C_j, C_{j+1}]$. So we obtain:

Level L1 contains the points $A_{ij}$ settled under $C_1$ and on $C_1$;

Level L2 contains the points $A_{ij}$ settled between $C_1$, $C_2$ and on $C_2$;

Level L3 the points between $C_2$ and $C_3$; and so on.

We call the **risk algorithm p.c.** (risk algorithm of parallel curves) all the above steps 1-9, together with the initial data introduction.

## 3. Numerical application based on the risk algorithm p. c.

*Remark 3.* In applications, the mathematical notations from section 2 have the following meanings: .

$m$ = the number of classes of probabilities; $m = 9$ on $Ox$ axis;

$n$ = the number of classes of impact; $n = 7$ on $Oy$ axis;

$r$ = the number of risk levels; $r = 6$;

$x_i = p_i$ the class of probability no. $i$;

$y_j = I_j$ the class of impact no. $j$;

$x_i = i$, $i = 1,9$; $y_j = j$, $j = 1,7$; $A_{ij} = A(P_i, I_j)$.

For simplicity we suppress the lower index and denote
$h_j = hj$, $C_j = Cj$, $R_j = Rj$, $B_j = Bj$.

So we use the notations:
$h2, h3, \cdots, h6$; $C1, C2, \cdots, C6$;
$R1, R2, \cdots, R6$; $B1, B2, \cdots, B7$.

By computations we obtain successively: $c = 9$
$(OV) \cap (xy = 9) \Rightarrow B1\left(9\sqrt{7}/7; \sqrt{7}\right)$
(normal in B1): $1.286x - y - 1.729 = 0$
(normal in B1) $\cap (y = 7) \Rightarrow B7(6.778; 7)$
$\|B1B7\| = 5.517$; $h = 5.517/6 = 0.919$
$h2 = h = 0.919$, $h3 = 2h = 1.838$
$h4 = 3h = 2.757$, $h5 = 4h = 3.676$
$h6 = 5h = 4.595$.

The curve Cj has the parametric equations

$$Xj(x, c, hj) = x + \frac{c \cdot hj}{\sqrt{c^2 + x^4}}, \quad j = 2, 6,$$

$$Yj(x,c,hj) = \frac{c}{x} + \frac{x^2 \cdot hj}{\sqrt{c^2 + x^4}}, \quad j = \overline{2,6}.$$

In order to draw the parallel curves $C_j$ we use the Mathcad library. The unit of measure on $Ox$ is different of the unit on $Oy$ axis.

In figure 2, on axis $Ox$ and $Oy$ is written respectively
$x, X2(x,c,h2), \cdots, X6(x,c,h6)$
$\frac{c}{x}, Y2(x,c,h2), \cdots, Y6(x,c,h6), 1, 2, \cdots, 6.$

The user must draw the parallel lines
$x = 1, x = 2, \cdots, x = 9$ in figure 2.

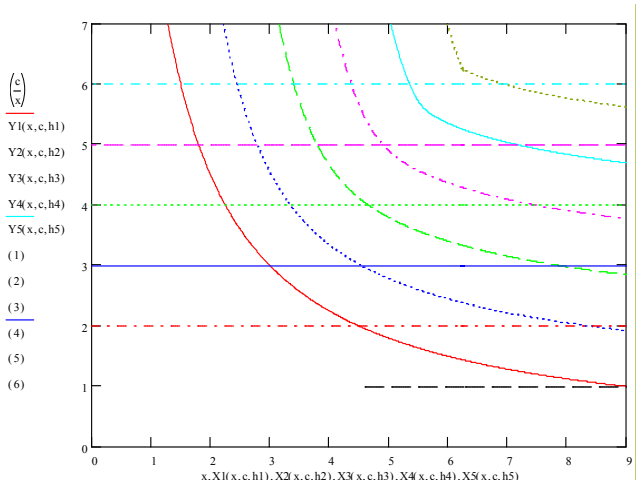

**Figure 2.** The six parallel curves of risk $C_j$.

The values of risks are obtained from 7.1 and 7.2 of algorithm:

$B1(9;1),\quad R1 = 9$
$B2(3.966; 3.372),\quad R2 = a_2 b_2 = 13.373$
$B3(4.531; 4.097),\quad R3 = a_3 b_3 = 18.564$
$B4(5.095; 4.823),\quad R4 = a_4 b_4 = 24.573$
$B5(5.659; 5.549),\quad R5 = a_5 b_5 = 31.402$
$B6(6.224; 6.274),\quad R6 = a_6 b_6 = 39.037.$

A number of 7 (seven) levels of risk were assessed according to the figure 2 to which correspond the following couples (**P**robability, **I**mpact):

Level 1: (1,1), (1,2), (1,3), (1,4), (1,5), (1,6), (1,7), (2,1), (2,2), (2,3), (2,4), (3,1), (3,2), (3,3), (4,1), (4,2), (5,1), (6,1), (7,1), (8,1), (9,1);
Level 2: (2,5), (2,6), (2,7), (3,4), (4,3), (5,2), (6,2), (7,2), (8,2);
Level 3: (3,5), (3,6), (3,7), (4,4), (5,3), (6,3), (7,3), (8,3), (9,2);
Level 4: (4,5), (4,6), (4,7), (5,4), (6,4), (7,4), (9,3);
Level 5: (5,5), (5,6), (5,7), (6,5), (7,5), (8,4), (9,4);
Level 6: (6,6), (6,7), (7,6), (8,5), (9,5);
Level 7: (7,7), (8,6), (8,7), (9,6), (9,7).

## 5. The generalization of the risk curve

Up to now we have used an explicit hyperbolic curve of risk $\gamma : xy = c$.

Now the curve $\gamma$ has the general explicit equation $y = f(x)$ with function $f$ continuous, derivable and $f'(x) \neq 0$ for let us say $x \geq 0$.

Let $M, N$ be two points
$M(x_0, y_0) \in \gamma$ and $N(X,Y) \in \Gamma$.

Like in the proposition 1 we use: slops, tangent equation, normal equation, translations and so on. So we obtain successively:

$m_{\tan} = f'(x_0), m_{norm} = -1/f'(x_0)$

$$y - y_0 = -\frac{1}{f'(x_0)}(x - x_0)$$

$X = x_0 + p, Y = y_0 + q$ ( real values $p$ and $q$ to be determined )

$$Y - y_0 = -\frac{1}{f'(x_0)}(X - x_0),$$

$q = -p/f'(x_0)$, $p^2 + q^2 = h^2$ ;we took

$$p = +h \frac{|f'(x_0)|}{\sqrt{1 + f'^2(x_0)}}$$

$$q = -h \frac{|f'(x_0)|}{f'(x_0)} \frac{1}{\sqrt{1 + f'^2(x_0)}}$$

$$X = x_0 + h \frac{|f'(x_0)|}{\sqrt{1 + f'^2(x_0)}} \quad (1^*)$$

$$Y = f(x_0) - h \frac{|f'(x_0)|}{f'(x_0)} \frac{1}{\sqrt{1 + f'^2(x_0)}}$$

We can suppress the index zero and the couple of equations (1*) is the generalization of (1).

If we put in (1*) the function $f(x) = c/x$, one obtains the parametric equations (1).

*Remark 4.* The risk algorithm p.c. with the steps 1-9 could be repeated for the function $f(x)$.

## 6. Other possibilities to define the measure of risk

In section 2 we have defined the measure of risk by using a hyperbolic curve and denoted $R_1 = x_m y_1$.

From mathematical point of view [6], chapters 2 and 3, the risk is a random variable $X$. In a finite state space we denote the discrete random variable

$$X = \begin{pmatrix} x_1 & x_2 & \cdots & x_i & \cdots & x_n \\ p_1 & p_2 & \cdots & p_i & \cdots & p_n \end{pmatrix}, \ p_i \geq 0, \sum_{i=1}^{n} p_i = 1$$

and the measure of risk is a real number $R(X)$.

There are several possibilities to define $R(X)$. In order to do that we recall several usual notations of probability domain :

$M(X) = m = \sum_{i=1}^{n} x_i p_i$ is the mean value;

$D^2(X) = \sigma^2(X) = Var(X) = M[(m-X)^2]$

is the dispersion or the variance of $X$ ;

$(x)_+$ means $(x)_+ = \max\{0; x\}$.

A non-shifted estimators of mean value and dispersion are

$$\overline{X} = \frac{1}{n} \sum_{i=1}^{n} x_i \ , \ s^2(X) = \frac{1}{n-1} \sum_{i=1}^{n} (\overline{X} - x_i)^2 \ .$$

Now we define several measures of risk.

**I.** $R_I(X) = Var(x)$ (7)

When the computation is based on estimators we denote it by $\overline{R}$

**II.** We denote by $T$ a threshold of risk and define

$$R_{II1}(X,T) = M[(T-X)_+]^2 \qquad (8)$$

(one eliminates the values $x_i > T$ )

$$R_{II2}(X,T) = M[(X-T)_+]^2 \qquad (9)$$

(one eliminates the values $x_i < T$ ).

**III**. Generalization

$$R_{IIIp}(X,T) = M[(T-X)_+]^p, \ p \in N^* \qquad (10)$$

**IV.** The risk measure based on **semi-variance** or **semi-dispersion**, denoted by S

$$SD^2(X) = S\sigma^2(X) = SVar(X)$$

$$R_{IV}(X) = SVar(X) = M[(m-X)_+]^2 \ (11)$$

An estimator of semi-dispersion is

$$Ss^2(X) = \frac{1}{n-1} \sum_{i=1}^{n} [(\overline{X} - x_i)_+]^2 \qquad (12)$$

The user is interested only in dispersion of values $x_i$ so that $x_i < \overline{X}$ .

An estimator of risk measure is

$\overline{R}_{IV}(X) = Ss^2(X)$ .

**V.** The Taguchi [6] measure of risk (in industrial practice) is defined as

$R_T(X,T) = k[Var(X) + (m-T)^2], k > 0$ (13)

An estimation of Taguchi's measure of risk is

$$\overline{R}_T(X,T) = k[s^2 + (\overline{X} - T)^2] \qquad (14)$$

Each user must choose the appropriate law of measuring the risk.

**Numerical examples.**

1. Let $X$ be a random variable with the probability repartition

$$X = \begin{pmatrix} 15 & 14 & 18 & 15 & 12 & 11 & 5 & 0 & 3 & 5 & 4 & 5 \\ \frac{1}{12} & \frac{1}{12} & \frac{1}{12} & \frac{1}{12} & \frac{1}{12} & \frac{1}{12} & \frac{1}{12} & \frac{1}{12} & \frac{1}{12} & \frac{1}{12} & \frac{1}{12} & \frac{1}{12} \end{pmatrix}$$

a). Risk measure based on estimators.

The mean value and its estimation are

$m = M(X) = 8.917; \overline{X} = 8.917$ .

The estimation of dispersion and risk measure (7) are

$s^2(X) = 32.354; \overline{R}_I(X) = 32.354$ .

b). The risk measure based on semi-dispersion is obtained as it follows

$$\sum_{i=1}^{12} [(\overline{X} - x_i)_+]^2 = 184.73$$

$\overline{R}_{IV}(X) = Ss^2(X) = 16.794$ .

*Remark 5* . Because we use only the estimators, the probabilities $p_i$ are not necessary

2. $X = (21 \ 21 \ 30 \ 22 \ 32 \ 19 \ 3 \ 3 \ 5 \ 8 \ 12 \ 11)$

a). The risk measure based on estimators.

$\overline{X} = 15.583; s^2(X) = 100.72$

$\overline{R}_I(X) = s^2(X) = 100.72$ .

b). The risk measure based on semi-dispersion

$$\sum_{i=1}^{12}\left[(\overline{X}-x_i)_+\right]^2 = 520.008$$

$$\overline{R}_{IV}(X) = Ss^2(X) = 47.273$$

## 7. Conclusions

The assessment of risks specific to a certain security system should be performed on strict and objective mathematic basis which can be repeated in time and the results obtained to be reproducible. The assessment of risks on particular levels, mathematically and objectively determined, allows selecting the options of treatment, so that the management decision should be impartial.

The proposed method has been elaborated for assessing the risk level specific to the information security within an organization, but it may be applied, with well results, in any risk assessment, depending on the probability of occurrence of an undesirable certain event generating the risk and consequences of the respective event.

In this purpose, the scales for the two axes (**P**robability and Impact) are conveniently selected and those two coordinates assessing the risk are represented on the graphic in figure 2.

According to the tolerable level accepted for the regarded risk, respectively to the plotted curve of risk, the positioning of the point in regard to this curve is to be analyzed. If the point is placed below the plotted curve, the presumed risk is acceptable and implicitly does not require for any decreasing measures to be adopted. Otherwise, if the point is placed upwards the risk curve, some measures have to be taken for decreasing the risk to a tolerable level.

The priorities of treatment are determined for each regarded risk according to the assessed risk level.

There are also others possibilities to define the value of risk, based on mean value and variance.

**References**

[1]. BS ISO/IEC 17799:2005: *Information technology – Security techniques – Code of practice for Information security management*.

[2]. BAICU Floarea, BAICU A.M., *Audit and Security of Information Systems,* Victor Publishing House, Bucharest, 2006.

[3]. BAICU Floarea , BAICU A.M., *Risk Management of Information Security Systems, complying with Deming cycle – PDCA,* review "Quality, access to success", year 8, nr 4, May 2007, ( to appear ).

[4]. BAICU Floarea , BAICU A.M., *Risks associated to Security of Information Systems,* Hyperion University Annals - Mathematics, Informatics 2006, pages 59-66.

[5]. POPOVICIU Nicolae , *Tensor Calculus. Theory and Applications*, Military Technical Academy of Bucharest Editing House, 2002.

[6]. RADULESCU M., RADULESCU S., RADULESCU C. Z., *Modele matematice pentru optimizarea investitiilor financiare*, Editura Academiei Române, 2006.